\documentclass[preprint,12pt, a4paper]{elsarticle} 

\usepackage{amssymb}

\usepackage[T1]{fontenc}
\usepackage{bigfoot} % to allow verbatim in footnote
\usepackage[numbered,framed]{matlab-prettifier}
\usepackage{filecontents}

\usepackage{lineno}
\usepackage{hyperref}

\lstMakeShortInline"

\begin{document}

\begin{frontmatter}



\title{MIPROT: A Medical Image Processing Toolbox for MATLAB}


\author{Alberto Gomez}

\address{School of Biomedical Engineering and Imaging Sciences, King's College London}

\begin{abstract}
This paper presents a Matlab toolbox to perform basic image processing and visualization tasks, particularly designed for medical image processing. The functionalities available are similar to basic functions found in other non-Matlab widely used libraries such as the Insight Toolkit (ITK). The toolbox is entirely written in native Matlab code, but is fast and flexible.

Main use cases for the toolbox are illustrated here, including image input/output, pre-processing,  filtering, image registration and visualisation. Both the code and sample data are made publicly available and open source.
\end{abstract}


\end{frontmatter}

%\linenumbers


\section{Introduction}

MATLAB\footnote{https://mathworks.com/products/matlab.html} (The Mathworks Inc.) is a widely used programming language and integrated development environment (IDE) with many engineering features, including image processing and computer vision. MATLAB is intuitive and easy to use, and allows for easy prototyping and rapid visualisation. As a result, many researchers use MATLAB for medical image processing. However, medical imaging presents a number of key differences with computer vision and natural image processing that require for specific processing tools. 

Although there is a large number of medical imaging work carried out in MATLAB, most researchers develop their own functionality. There is also a large amount of software made available to the community to address some specific aspects of medical image processing, such as image input/output \cite{DJKroon2011}, image segmentation \cite{KeiOtsuka2020}, feature extraction of medical images \cite{liebgott2018imfeatbox}, and graphical user interfaces for research pipelines \cite{brossard2020mp3}.

However, the vast majority of medical imaging algorithmic research and development is implemented in C++ and python using specific libraries such as the Insight Toolkit\footnote{https://itk.org/} (ITK), MIRTK\footnote{https://mirtk.github.io/}, and more recently libraries that integrate machine learning and deep learning features such as MONAI\footnote{https://github.com/Project-MONAI/MONAI/}. The lack of a unified framework to manage, analyse, process and visualise medical images in MATLAB, that facilitates implementation of common techniques such as segmentation, registration and classification motivated the work in this paper.

This paper describes the Medical Image Processing Toolbox for MATLAB (MIPROT), an open source, free software aiming at facilitating research and development with medical images. The software is available from the \href{https://uk.mathworks.com/matlabcentral/fileexchange/41594-medical-image-processing-toolbox}{MATLAB Central File Exchange}\footnote{\href{https://uk.mathworks.com/matlabcentral/fileexchange/41594-medical-image-processing-toolbox}{mathworks.com/matlabcentral/fileexchange/41594-medical-image-processing-toolbox}}, and the most up-to-date version can be downloaded from the public \href{https://gitlab.com/compounding/matlab}{GitLab} repository\footnote{\href{https://gitlab.com/compounding/matlab}{https://gitlab.com/compounding/matlab}}.

This paper is organised as follows. First, the specific aspects of medical images that need to be covered are described in Section \ref{sec:medical-images}. Section \ref{sec:overview} gives a description of the software, focusing on image I/O, basic functionalities, filtering and visualisation. overview of the software. Section \ref{sec:examples} provides some example applications including image registration and image segmentation.

\section{Specific Aspects of Medical Images}
\label{sec:medical-images}

Medical imaging requires to consider particularities of medical images that are normally disregarded in natural images. Not only the physics of the acquisition systems (magnetic resonance -MR, ultrasound -US, computed tomography -CT, etc) are different to optical systems employed in natural images, but also medical images are embedded in a coordinate system where, beyond image intensity, pixel size and image axes bear important information about the anatomy. Moreover, the recorded intensity may have a physiological meaning. As a result, especial care must be taken when performing basic image processing operations such as re-sizing, re-sampling, or intensity re-scaling, so as to consider the impact of these operations on the medical information contained in the image.

%Explain why the software is important, and describe the exact (scientific) problem(s) it solves.

For this reasons, dedicated medical image processing libraries have been developed for some programming languages, such as Python or C++, with the most popular example being the Insight Toolkit (ITK) \cite{yoo2002engineering}. However, although MATLAB is very widely used for both computer vision and medical imaging, and there is a large collection of built-in functions and classes for computer vision within, there is a lack of a structured toolbox for medical image processing. To this end, this paper describes a MATLAB toolbox for medical image processing. The toolbox is focused around basic 2D, 3D and 4D medical image processing tasks, including image input/output (I/O), spatial transforms, cropping, resizing, slicing, and visualization. The toolbox also provides basic mesh I/O. %It also includes integration of more advanced computer vision tools with medical images. 

\section{Organisation of the Software}
\label{sec:overview}

%Describe the software in as much as is necessary to establish a vocabulary needed to explain its impact. 

\begin{figure}[!htb]
    \centering
    \includegraphics[width=\linewidth]{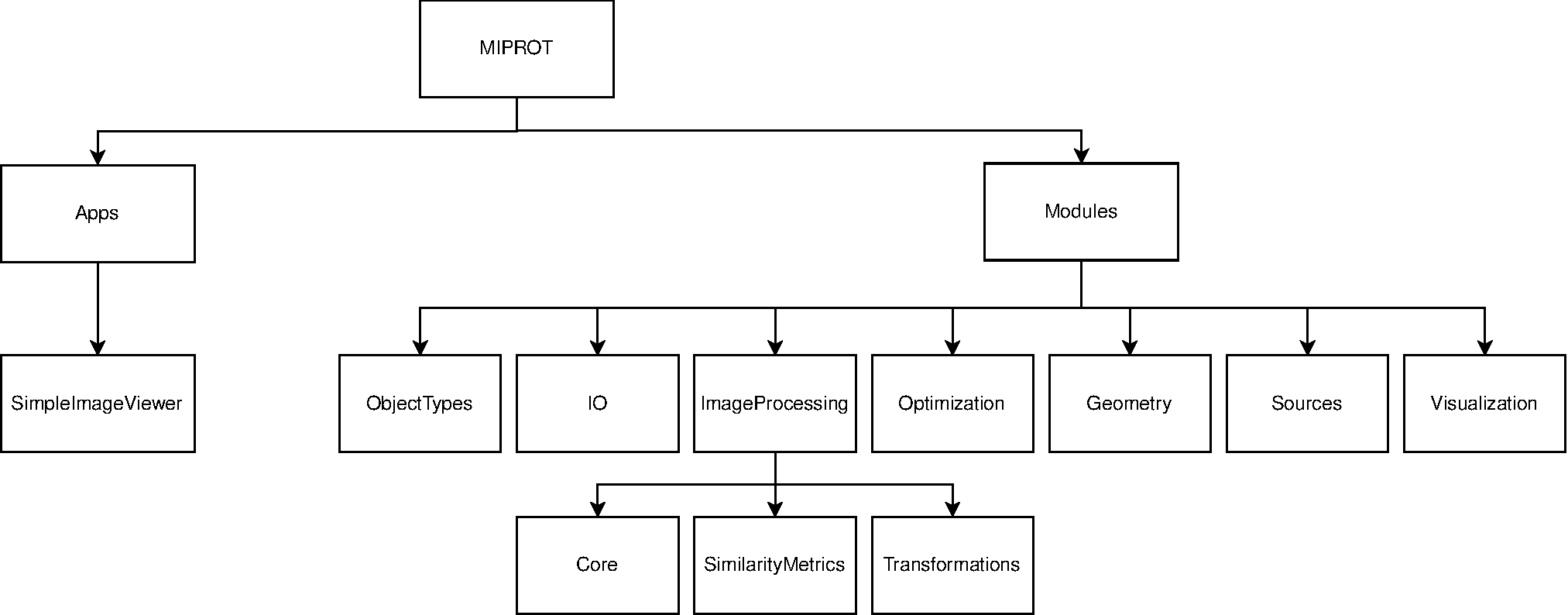}
    \caption{Overview of MIPROT organisation.}
    \label{fig:softwareoverview}
\end{figure}

This toolbox is built around the class \texttt{ImageType}, which represents $N$-dimensional medical images, and heavily inspired by the \texttt{itk::Image} class from ITK. Figure \ref{fig:softwareoverview} shows the organisation in modules, and the functionality is described in turn over the following subsections. Section \ref{sec:simpleviewer} describes an app that features a graphical interface to visualize images interactively.

\subsection{Module `Object Types'}

\begin{figure}[!htb]
    \centering
    \includegraphics[width=\linewidth]{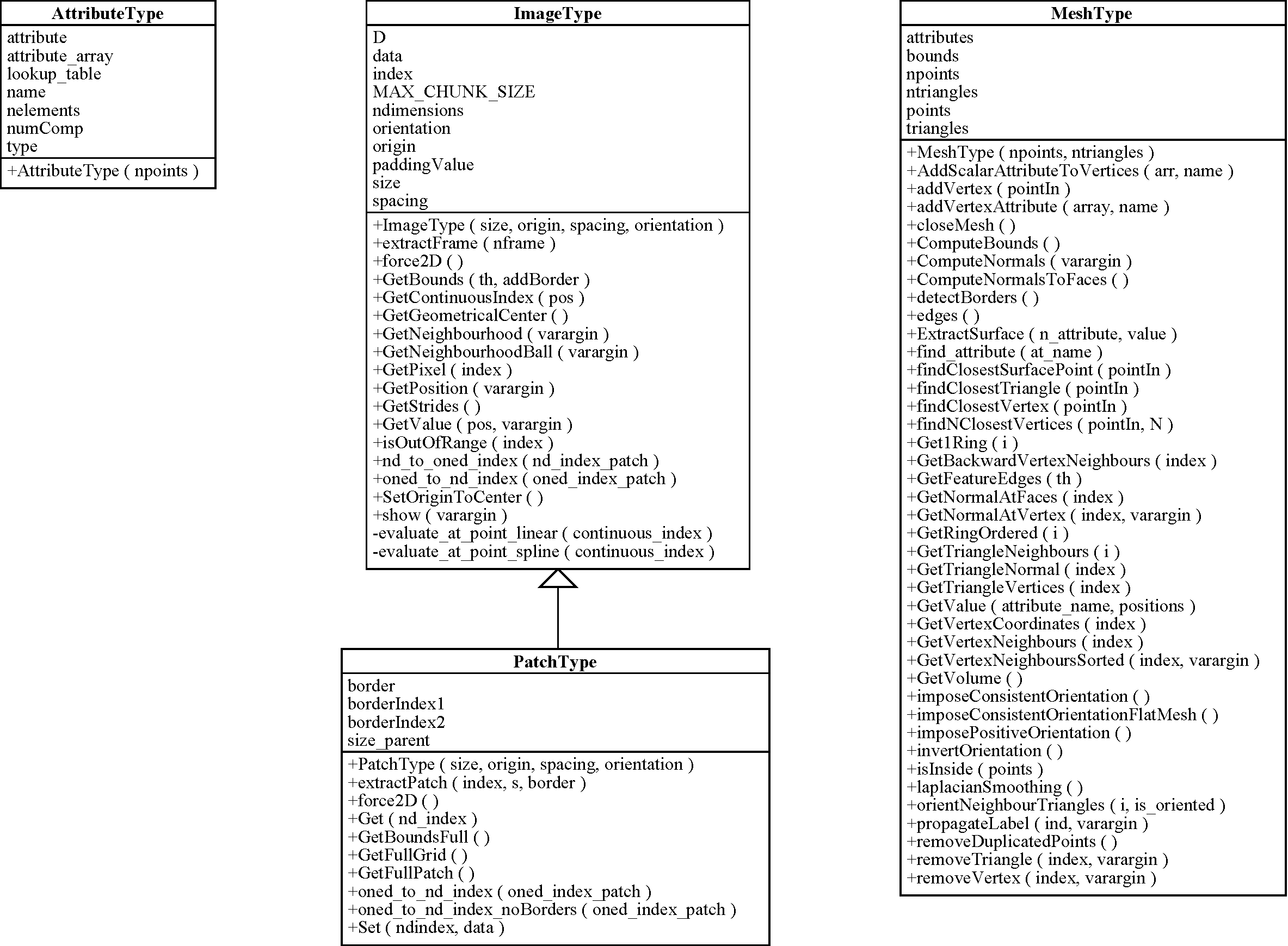}
    \caption{Main object types: \texttt{ImageType}, \texttt{PatchType}, \texttt{MeshType} and \texttt{AttributeType}.}
    \label{fig:imagetype}
\end{figure}

The main object types are illustrated in Fig. \ref{fig:imagetype}. \texttt{ImageType} represents an image, and its subclass \texttt{PatchType} can be used to represent a patch within an image. \texttt{MeshType} represents triangulated meshes (for example, for image segmentation), and each node of the mesh can contain one or more attributes of type \texttt{AttributeType}.

\noindent{\textbf{ImageType}: } The \texttt{ImageType} is a container that allows to store a $N$-dimensional image (currently 2D, 3D and 4D are supported). The class contains the image pixel values (stored in the \texttt{.data} member as a double matrix), and the geometry defined by the \texttt{.origin} member ($N \times 1$ matrix with the  world coordinates of the pixel with index $(1,...)$), the \texttt{.spacing} member ($N \times 1$ matrix with the pixel size), and the \texttt{.orientation} member (a $N \times N$ matrix defining the axis of the image, identity by default). The image class also features the following members:
\begin{itemize}
    \item \texttt{.index} The index of the first pixel. By default this is $\mathbf{1} \in \mathbb{N}^N$. However, this could change and particularly for images of type \texttt{PatchType} as described later.
    \item \texttt{.paddingValue} is by default 0, and is a value that will be used for padding in operations that require values outside the image extent, for example after a transformation.
    \item \texttt{.size} contains the size of the data matrix.
    \item \texttt{.MAX\_CHUNK\_SIZE} is a number defining the maximum number of pixels to be processed at once by filters that operate pixel-wise. This allows to limit memory usage of the methods at the cost of longer processing time.
\end{itemize}
To construct a new image, the user can define an empty image from the geometry or use an existing image as template (all members are copied):
\begin{lstlisting}[caption = {Creating an image}]
sz = [100 120 130]'; % size, in numer of voxels
sp = [1.1 1.0 0.9]'; % spacing, in mm
or = -(sz-1)/2.*sp; % origin, with point (0,0,0) at centre
im = ImageType(sz, or, sp); % image created

im2 = ImageType(im); % copy all members from template
\end{lstlisting}

In addition to these members, \texttt{ImageType} features the following member functions (arguments marked with `<>' are optional:
\begin{itemize}
    \item \texttt{.extractFrame($n$)} From a 4D image extracts the $n$-th 3D frame.
    \item \texttt{.force2D()} Remove all singleton dimensions to make the image 2D, if the input is 3D.
    \item \texttt{.GetBounds($th$)} obtains the $2N \times 1$ vector with the min and max coordinate values beyond which the pixel values are below $th$.
    \item \texttt{.GetContinuousIndex($pos$)} extract the index, as a double, corresponding to the position $pos$.
    \item \texttt{.GetGeometricCentre()} calculates the world coordinates of the center of the image.
    \item \texttt{.GetNeighbourhood(<$n$>} Get the relative indices of the fully connected neighbourhood of width $n$. By default, $n=1$. This is convenient for neighbourhood based operators. The shape of this neighbourhood is an $N$-dimensional cube.
    \item \texttt{.GetNeighbourhoodBall(<$n$>} Get the relative indices of the fully connected neighbourhood ball of width $n$. By default, $n=1$. This is convenient for neighbourhood based operators. The shape of this neighbourhood is an $N$-dimensional sphere.
    \item \texttt{.GetPixel($idx$)} get the intensity value at pixel with index $idx$. Indices start at 1.
    \item \texttt{.GetPosition(<$cidx$>)} returns the position (or positions) in world coordinates of pixels with indices $cidx$, which can be a continuous index. $cidx$ should be a $N x m$ matrix, where $m$ is the number of indices. if no argument is passed, the function returns all positions of the image. If $N=1$ and the image dimension is $>1$, then the indices are interpreted as linear indices. This allows to do things like the following, to extract the positions of the pixels where values are above a threshold:
    \begin{lstlisting}[caption = {Creating an image}]
    th = 0.5;
    indices = find(im.data > th);
    positions = im.GetPosition(indices);
    \end{lstlisting}
    \item \texttt{.GetStrides()} Returns the strides to compute linear indices.
    \item \texttt{.GetValue($pos$, $<mode>$} Interpolates the image intensity value at position(s) $pos$. The interpolation $mode$ is optional (by default is `NN') and can take the following values, as strings: 'NN' (nearest neighbour), 'linear' or 'spline'.
    \item \texttt{.isOutOfRange($idx$)} returns \texttt{true} if the index is outside the image extent.
    \item \texttt{.nd\_to\_oned\_index($ndidx$)} and \texttt{.oned\_to\_nd\_index($idx$)} convert from $N$d to linear indices and vice-versa.
    \item \texttt{.SetOriginToCenter())} is a convenience function to center the image.
    \item \texttt{.show(<options...>))} is a simple visualization function. 2D and 3D images are supported. For 2D images it shows the image in grayscale. For 3D images, a 3D representation of 3 orthogonal planes is shown. The following options (and option pairs) allow further customization:
    \begin{itemize}
        \item `opacity', opacityval : sets the opacity of the image to a value between 0 and 1.
        \item `colorrange', range : a tuple with the min and max value of colors to show. The image range is stretched to that range.
        \item `colormap', cm: sets the colormap from the default (gray) to any other.
        \item `mpr': shows 3D images as a 2D multi-planar-reformatting view.
        \item `mproffset': set the coordinates of the point where the mpr planes meet. By default, the centre of the image.
    \end{itemize}
\end{itemize}

\noindent{\textbf{PatchType}: } The \texttt{PatchType} is a subclass of \texttt{ImageType} that is used internally by some algorithms and for most everyday processing can be safely ignored.

\noindent{\textbf{MeshType}: } The \texttt{MeshType} class contains triangular meshes. The main members are the \texttt{.triangles}, which contain a $3 \times n$ matrix with one column per triangle, and each row represents the index of the vertex; and \texttt{.points} which contain the list of vertices coordinates. In addition, the mesh has the \texttt{.attributes} member which is an array of one or more attributes \texttt{AttributeType}. Each attribute can be associated to the vertices or to the triangles. Association is established depending on the number of elements.

\subsection{Module `IO'}

This module provides functions to read and write images and meshes. The functions provided are:
\begin{itemize}
    \item \texttt{read\_gipl(filename)} Read a *.gipl image.
    \item \texttt{write\_gipl(filename, im)} Writes the image \texttt{im} to the *.gipl file.
    \item \texttt{read\_mhd(filename)} Read a *.mhd/*.raw image.
    \item \texttt{write\_mhd(filename, im <, options>)}  Writes the image \texttt{im} to *.mhd/*.raw files (only the *.mhd needs to be passed as argument). the options are:
    \begin{itemize}
        \item \texttt{`elementtype', type}: choose the pixel type, \texttt{type}=\{\texttt{`uint8'}, ...\}
    \end{itemize}
    \item \texttt{read\_nifty(filename)} Read a *.nii image. Writing is not yet supported.
    \item \texttt{read\_picture(filename)} Read a *.png, jpeg, tiff, etc. images.
    \item \texttt{read\_MITKPoints(filename)} Read an *.mps MITK file.
    \item \texttt{read\_MITKPointsFast(filename)} Read an *.mps MITK file (faster implementation)
    \item \texttt{write\_MITKPoints(filename, pts)} Write a point list to an *.mps MITK file.
    \item \texttt{read\_stlMesh(filename)} Read *.stl meshes.
    \item \texttt{write\_stlMesh(filename, mesh)}Write to *.stl file.
    \item \texttt{read\_vtkMesh(filename)} Read text (legacy) *.vtk mesh files. 
    \item \texttt{write\_vtkMesh(filename, mesh)} Write to text (legacy) *.vtk mesh files. 
    \item \texttt{write\_deformetricaVTKShape(filename, mesh)} Write mesh to a deformetrica file.
    \item \texttt{read\_ITKMatrix(filename)} Read a $4 \times 4$ matrix (which can be used as a 3D affine matrix) from a text file. The file format should be with a first line commented with `\#' followed by 4 lines of space separated values, as follows:
    \begin{verbatim}
    # itkMatrix 4 x 4
    1.0 0.0 0.0 0.0
    0.0 1.0 0.0 0.0
    0.0 0.0 1.0 0.0
    0.0 0.0 0.0 1.0
    \end{verbatim}
    \item \texttt{write\_ITKMatrix(filename, M)} writes a matrix M to the file `filename'. \texttt{*.mat} extension is preferred.
\end{itemize}
As an example, one can read an image and write it into a different format:
 \begin{lstlisting}[caption = {Image I/O. The image is read in gipl, and regardless of the pixel format, it is saved as unsigned char.}]
filename_in = 'myfolder/image.gipl';
im = read_gipl(filename_in);
% convert the range to [0,255] to save as uint8
im.data = im.data / max(im.data(:)) * 255; 
filename_out = 'myfolder/image.mhd';
write_mhd(filename_out, im, 'ElementType','uint8');
\end{lstlisting}

\subsection{Module `ImageProcessing'}

The image processing module has three submodules: \texttt{Core}, \texttt{SimilarityMetrics} and \texttt{Transformations}

\subsubsection{Core}
\label{sec:core}
The \texttt{Core} module includes basic operations:
\begin{itemize}
    \item \texttt{cropImage(image, bounds)} Crops the image to the bounds specified by the $2N \times 1$ vector, which should contain the min and max values along each dimension (in world coordinates).
    \item \texttt{gradientImage(image <, options>)} Computes the gradient of an image, and returns a $N+1$ list with the gradient over each of the $N$ dimensions. Features the following options:
    \begin{itemize}
        \item \texttt{`order', o}: applies the derivative of order \texttt{o}. By default, $o=1$. 
    \end{itemize}
    \item \texttt{resampleImage(image, ref <, options> )} Resamples the input \texttt{image} to the grid defined by the \texttt{ref} image. The function can also be used without a reference image, using the options as follows:
    \begin{itemize}
        \item Resample to a specific spacing
        \begin{lstlisting}
% downsample the image to pixels twice as big:
im2 =resampleImage(im, [], 'spacing', im.spacing * 2);
        \end{lstlisting}
        \item Resample the image to a certain spacing and size, by using padding/cropping under the hood if needed:
        \begin{lstlisting}
% resample to a spacing of [1, 1] and a size of [100, 100]
im2 = resampleImage(im, [], 'spacing_and_size', ...
                    [1 1 100 100]');
        \end{lstlisting}
        \item Resample the image to a certain spacing, size and centre, by using padding/cropping under the hood if needed:
        \begin{lstlisting}
% resample to a spacing of [1, 1] and a size of [100, 100], and center at [0 0].
im2 = resampleImage(im, [], ...
                    'spacing_and_size_and_centre', ...
                    [1 1 100 100 0 0]');
        \end{lstlisting}
    \end{itemize}
    This function also allows to define the interpolation by adding the option \texttt{`interpolation', `NN'},  or \texttt{`iterpolation', `linear'} to the arguments.
    \item \texttt{resliceImage(image <, options>)} This function allows to slice a 3D image to obtain an oblique slice. Returns a 3D image with the oriented slice, and a 2D image. The options to define the slice are indicated by the following examples:
    \begin{lstlisting}
% reslice defining the slicing plane with a 4x4 matrix
M = eye(4); % this will extract the central XY plane.
[slice, slice2D] = resliceIMage(im, 'mat', M);
    \end{lstlisting}
    \begin{lstlisting}
% reslice defining the slicing plane with a point and a normal. Equivalent to the previous example.
n = [0 0 1]';
point = [0 0 0]';
[slice, slice2D] = resliceIMage(im, 'plane', n, point);
    \end{lstlisting}
    \begin{lstlisting}
% reslice and set the spacing of the slice (otherwise average spacing is used)
M = eye(4); % this will extract the central XY plane.
[slice, slice2D] = resliceIMage(im, 'mat', M, 'spacing', [0.5 0.5]');
    \end{lstlisting}
    \begin{lstlisting}
% reslice and take 3 continguous slices (the resulting image will be 3D)
n = 3;
[sl, sl] = resliceIMage(im, 'mat', M, 'thicknes', n);
    \end{lstlisting}
\end{itemize}

\subsubsection{SimilarityMetrics}
This module includes implementation of the normalised cross-correlation (NCC), normalised mutual information (NMI) and sum of squared differences (SSD) commonly used in image registration, as exemplified in Section \ref{sec:registration}.

\subsubsection{Transformations}

This module includes geometric transformation models that are typically used in image registration. The main functions are:
\begin{itemize}
    \item \texttt{transformRigid} applies an affine transform to an image.
    \item \texttt{transformFFDSplines} applies a free-form deformation to an image, using a BSpline grid.
\end{itemize}
And helper functions to obtain parameters from affine/rigid matrices and vice-versa. These will be exemplified in Section \ref{sec:registration}.

\subsection{Module `Optimization'}

This module includes commonly used optimization functions, including RANSAC \cite{fischler1981random}, the  Broyden-Fletcher-Goldfarb-Shannon method and variants (BFGS, implemented by D.Kroon from the University of Twente, 2010) and the conjugate gradient method (implemented by Ian T Nabney). Examples of how they can be used can be found in Sec \ref{sec:registration}.

\subsection{Module `Geometry'}

The \texttt{Geometry} module includes auxiliary low-level functions that are used by other functions to do basic geometric calculations. These functions are not intended to be used by the toolbox user, and are aimed at developers.

\subsection{Module `Sources'}

\begin{figure}[!htb]
    \centering
    \includegraphics[width=\linewidth]{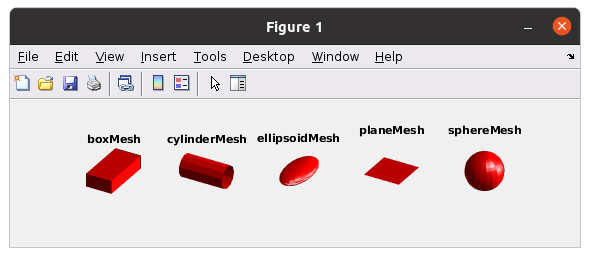}
    \caption{Mesh sources from the \texttt{Sources} module.}
    \label{fig:sources}
\end{figure}

The \texttt{Sources} module includes mesh and image sources, in the style of Paraview sources. Examples of mesh sources are indicated below and the result is shown in Fig. \ref{fig:sources}.
\begin{lstlisting}[caption = {Demonstration of the \texttt{sources} module.}]
% box
center = [0 0 0]';
dims = [10 20 5]';
b = boxMesh(center, dims);
subplot(1, 5, 1)
viewMesh(b);
view(30, 30);
axis equal
title('boxMesh')

% cylinder
ax = [1 0 0]';
radius = 5;
height = 20;
resolution = 12;
b = cylinderMesh(ax,center,radius, height,resolution);
subplot(1, 5, 2)
viewMesh(b);
view(30, 30);
axis equal
title('cylinderMesh')

% ellipsoid
radius = [10 20 5]';
b = ellipsoidMesh(center,radius, 'resolution', 20);
subplot(1, 5, 3)
viewMesh(b);
view(30, 30);
axis equal
title('ellipsoidMesh')


% plane
point = [2 0 0]'; 
normalVector = [0 0 1]';
b = planeMesh(point, normalVector, 'scale', 2);
subplot(1, 5, 4)
viewMesh(b);
view(30, 30);
axis equal
title('planeMesh')

% sphere
radius = 10;
b = sphereMesh(center,radius, 'resolution', 20);
subplot(1, 5, 5)
viewMesh(b);
view(30, 30);
axis equal
title('sphereMesh')
\end{lstlisting}

\subsection{Module `Visualization'}

The \texttt{Visualization} module includes auxiliary visualization functions to display meshes, 2D and 3D points, bounding boxes and axes representations:
\begin{itemize}
    \item \texttt{plotAxis(direction <, options>)}: display 3 colored arrows with the x, y and z axes defined by the $4 \times 4$ matrix \texttt{direction}. The matrix defines also the origin where the axes meet on the fourth column.
    \item \texttt{plotboundingbox(bounds <, options>)} displays a wireframe box from the $2N \times 1$ bounds vector.
    \item \texttt{plotpoints()} and \texttt{plotpoints2()} are wrappers of the \texttt{plot} function to plot respectively 3D and 2D points from a $3 \times n$ and a $2 \times n$ array. After the matrix, they take the same optional arguments as the \texttt{plot} function.
    \item \texttt{viewMesh(m <, options>)} displays a \texttt{MeshType} object, as demosntrated in Fig. \ref{fig:sources}. It takes the following optional arguments:
    \begin{itemize}
        \item \texttt{`showvectors', scale, attributen}: displays (alongside the mesh) the vectors in the mesh attribute number \texttt{attributen} scaled by the factor \texttt{scale}.
        \item \texttt{`labelColor', attributen}: uses the scalar values in the attribute number \texttt{attributen} to color the mesh. Otherwise all mesh is displayed with the same color.
        \item \texttt{`triangles', trianglelist}. Only display the triangles in \texttt{trianglelist}.
        \item \texttt{`axes', a}. Visualize the mesh on the figure axes \texttt{a}.
        \item \texttt{`wireframe'}: Display the mesh as a wireframe.
        \item \texttt{`featureedgescolor', color}: display the feature edges (edges which are a sharp ridge in the surface) with the color \texttt{color}, which should be an [r g b] array.
        \item \texttt{`wireframeSurface'}: display both the surface and the edges as wireframe.
        \item \texttt{`color', c}: use the color \texttt{c} for the surface.
        \item \texttt{`Tag', t}: Assign the tag \texttt{tag}.
        \item \texttt{`opacity', o}: apply the opacity \texttt{o} $\in [0, 1]$ to the surface.
    \end{itemize}
\end{itemize}

\section{Example Applications}
\label{sec:examples}

In this section we present a number of examples of everyday medical image processing tasks to illustrate how the toolbox may be used.

\subsection{Image Preprocessing}

The toolbox includes a collection of functions for convenient pre-processing of images that can be used prior to other tasks. Here we illustrate this capabilities with some examples.

\begin{itemize}
    \item Image resampling to change image resolution, as shown in the examples in Sec. \ref{sec:core}.

    \item Image cropping to focus on a region of interest.
    \begin{lstlisting}
im = read_picture([folder '/mri1.jpg']);
bounds = [350 650 400 550]';
roi = cropImage(im, bounds);

subplot(1,2,1)
hold on;
im.show()
plotboundingbox(bounds, 'LineWidth',3); 
hold off;
axis equal;
axis off; 
title('Input image')

subplot(1,2,2)
roi.show();
axis equal;
axis off;
title('RoI')
    \end{lstlisting}
    \begin{figure}[!htb]
    \centering
    \includegraphics[width=\linewidth]{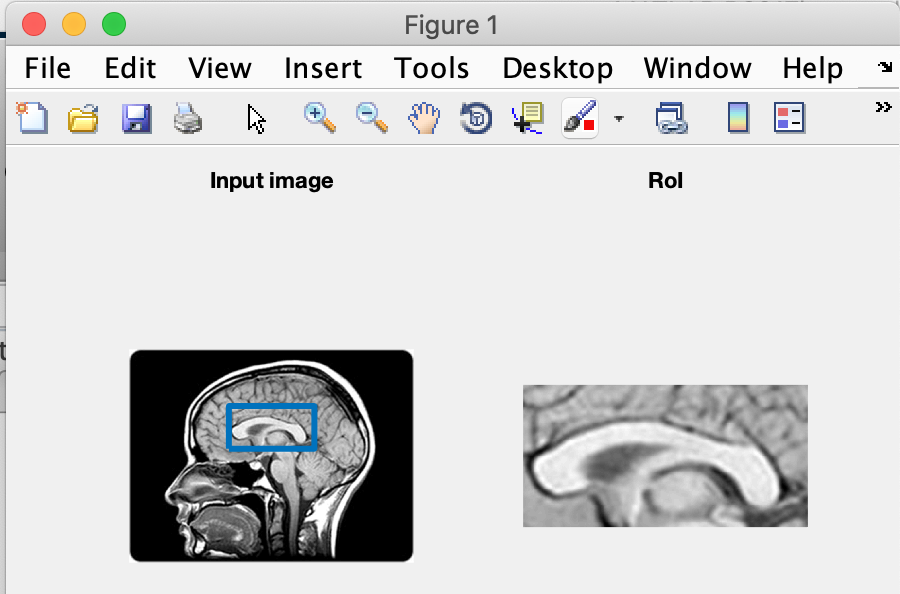}
    \caption{Cropping an image with a bounding box.}
    \label{fig:reg_rigid_2D}
\end{figure}
    \item Blurring and computing image gradients. In this example we first downsample the input image, and then we blur it by resample it to its own grid and adding a Gaussian blurring. Then we compute the derivatives along x and y,  using the derivative of a Gaussian.
    \begin{lstlisting}
im = read_mhd( 'data/CT1.mhd');
im = resampleImage(im, [], 'spacing', [2 2]');
neigh = [5 5]'; sigma = [3 3]';
im_blurred = resampleImage(im, im, 'blur', neigh, sigma);

gradients1 = gradientImage(im);
gx = ImageType(im);
gx.data = gradients1(:,:,1);
gy = ImageType(im);
gy.data = gradients1(:,:,2);

subplot(1,4,1)
im.show(); axis equal; axis off; 
title('Input')
%
subplot(2,4,2)
im_blurred.show(); axis equal; axis off; 
title('Blurred')
%
subplot(2,4,3)
gx.show(); axis equal; axis off; 
title('Gx')

subplot(2,4,4)
gy.show(); axis equal; axis off; 
title('Gy')
    \end{lstlisting}
     \begin{figure}[!htb]
    \centering
    \includegraphics[width=\linewidth]{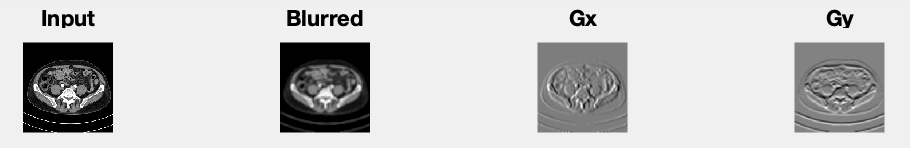}
    \caption{Blurring and computing derivatives.}
    \label{fig:reg_rigid_2D}
\end{figure}
\end{itemize}

\subsection{Rigid Image Registration}
\label{sec:registration}

In this example we demonstrate 2D image registration between 2 MRI images. The data was borrowed from the Elastix \cite{klein2009elastix} public repository. The results of the registration are illustrated in Fig. \ref{fig:reg_rigid_2D}. This example uses normalised cross-correlation (\texttt{similarityMetric\_NCC}) as similarity measure and the \texttt{fminlbfgs} optimizer.

\begin{lstlisting}[caption = {Example of rigid registration.}]
fixed = read_mhd(['data/fixed.mhd']);
moving= read_mhd(['data/moving.mhd']);

%% Add some random displacement to the moving image
moving=transform_rigid(moving, [0 0 15*pi/180], 'ref', moving);

%% Display images initially
difference = ImageType(fixed);
difference.data = fixed.data - moving.data;
figure;
subplot(2,3,1); 
fixed.show(); axis equal; axis off; title('Fixed')
subplot(2,3,2); 
moving.show(); axis equal; axis off; title('Moving')
subplot(2,3,3); 
difference.show(); axis equal; axis off; title('Difference before reg.')

%% Set up the registration process.
x0 = [0 0 0];

options = optimset('GradObj','off', 'Display', 'iter');
transform_object.transform_function = @(moving, params) transform_rigid(moving, params, 'ref', fixed);

f=@(x) similarityMetric_NCC( moving,fixed,transform_object,x);

%% Run registration

X = fminlbfgs(f, x0, options);

moving_tx = transform_rigid(moving, X, 'ref', fixed);
difference.data = fixed.data - moving_tx.data;
subplot(2,3,5); moving_tx.show(); axis equal; axis off; title('Moving after reg')
subplot(2,3,6); difference.show(); axis equal; axis off; title('Difference after reg.')
\end{lstlisting}

The example script displays the images before and after registration. The moving image is initially rotated 15 degrees, and the optimization recovers $-14.95$ degrees, which is an accuracy of 99.67\%.

\begin{figure}[!htb]
    \centering
    \includegraphics[width=\linewidth]{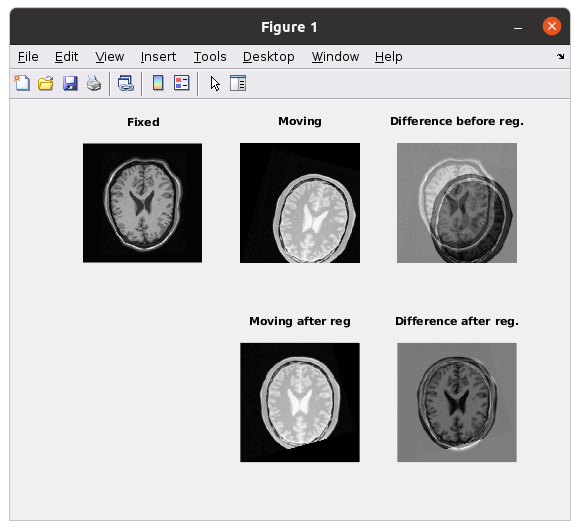}
    \caption{Result of rigid registration between two 2D images.}
    \label{fig:reg_rigid_2D}
\end{figure}

%\subsection{Rigid Image Registration with masks}
%In this example we demonstrate the use of masks to only use some region of the images during the registration process. This is a generic and flexible way of using padding (ignoring certain pixels) without the need to modify the input image.

\subsection{Deformable Image Registration}

In this example we demonstrate image registration using B-spline based free form deformations. This is a simple example with a single level B-spline grid, using B-splines of order 1 (linear). More information about B-spline based registration can be found in \cite{rueckert1999nonrigid}. 

\begin{lstlisting}[caption = {Example of deformable B-spline registration.}]
fixed = read_picture([getenv('HOME') 'data/c1.png']);
moving= read_picture([getenv('HOME') 'data/c3.png']);

% Downsample data
fixed = resampleImage(fixed, [], 'spacing',fixed.spacing.*[3 4]');
moving= resampleImage(moving, [], 'spacing',moving.spacing.*[3 4]');

%% Display images initially
difference = ImageType(fixed);
difference.data = fixed.data - moving.data;
figure;
subplot(2,3,1); fixed.show(); axis equal; axis off; title('Fixed')
subplot(2,3,2); moving.show(); axis equal; axis off; title('Moving')
subplot(2,3,3); difference.show(); axis equal; axis off; title('Difference before reg.')

%% Set up the registration process.
transform_params.nlevels=1;
transform_params.bsd=1;
transform_params.bounds=-1; % if -1 then take from image
transform_params.grid_spacing = [10 10]';

transform_initialization = transform_FFDBsplines_initialize(  moving, transform_params);

x0 = randn(size(transform_initialization.params0{1}))*10;

options = optimset('GradObj','off', 'Display', 'iter', 'MaxIter', 100);
transform_object.transform_function = @(moving, params) transform_FFDBsplines(moving, params, 'init', transform_initialization);

f = @(x) similarityMetric_NCC( moving, fixed, transform_object, x);

%% Run registration
X = fminlbfgs(f, x0, options);

moving_tx = transform_FFDBsplines(moving, X, 'init', transform_initialization);
difference.data = fixed.data - moving_tx.data;
subplot(2,3,5); moving_tx.show(); axis equal; axis off; title('Moving after reg')
subplot(2,3,6); difference.show(); axis equal; axis off; title('Difference after reg.')
\end{lstlisting}

\begin{figure}[!htb]
    \centering
    \includegraphics[width=\linewidth]{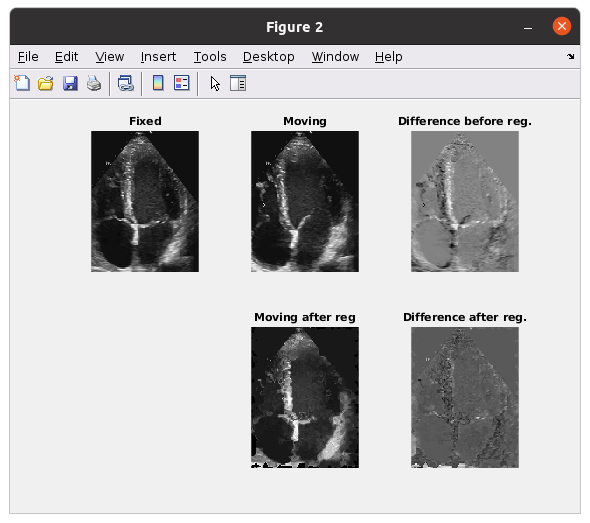}
    \caption{Result of non-rigid registration between two 2D ultrasound images of the heart, acquired at different phases of the cardiac cycle.}
    \label{fig:reg_rigid_2D}
\end{figure}

\subsection{Other applications}

Since the software was initially released in 2012, the toolbox has been intensively used by different researchers. Some examples of papers where this toolbox has been used for image preprocessing or visualization include: extending the viewer for the Matlab version of the flow profile extraction software presented in \cite{gomez2018optimal}; Radiotheraphy image analysis \cite{niko2014calculation}; classification of cardiac disease \cite{marciniak2017automatic}; manual registration of ultrasound images \cite{gomez2017fast}; and many others.

\section{Simple Image Viewer}

\begin{figure}[!htb]
    \centering
    \includegraphics[width=\linewidth]{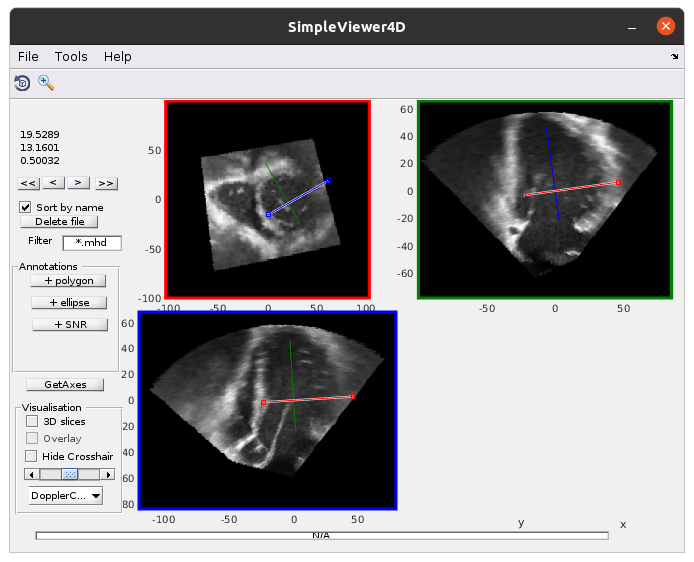}
    \caption{Main interface of the simple viewer showing a 3D ultrasound image of the heart.}
    \label{fig:simpleviewer}
\end{figure}

The toolbox also includes an image viewer graphical user interface, named \texttt{SimpleViewer\_GUI}. The main interface is illustrated in Fig. \ref{fig:simpleviewer}.

The viewer allows to load 2D, 3D and 3D+t images, and select oblique slices by tilting the cross-hairs on each panel, as shown in the figure. Other functionalities include:
\begin{itemize}
    \item Obtain the current slicing axes as a $4 \times 4$ matrix.
    \item Overlay two images with two different colormaps.
    \item Switch colormaps.
    \item Manual 2D segmentations (using polygon and ellipse tool).
    \item Manual rigid registration.
\end{itemize}

\section{Conclusion}

We have presented the Medical Imaging Processing Toolbox for Matlab. This toolbox provides flexibility and easy of use for everyday medical image analysis tasks and can be easily extended and integrated into other pipelines.

\section*{Acknowledgements}
\label{}

This work was partly supported by the Wellcome Trust UK (110179/Z/15/Z, 203905/Z/16/Z). A. Gomez also acknowledge financial support from the Department of Health via the National Institute for Health Research (NIHR) comprehensive Biomedical Research Centre award to Guy's and St Thomas' NHS Foundation Trust in partnership with King's College London and King's College Hospital NHS Foundation Trust.



\section*{References}
%% References:
%\bibliographystyle{elsarticle-num} 
%\bibliography{bibliography}


\end{document}